\documentclass[%
 reprint,
 amsmath,amssymb,
 aps,
aas_macros]{revtex4-2}

\usepackage{graphicx}% Include figure files
\usepackage{dcolumn}% Align table columns on decimal point
\usepackage{bm}% bold math
\usepackage{ulem}[normalem]
\usepackage{xcolor}
\usepackage{amssymb}
\usepackage{aas_macros}
\usepackage{hyperref}
\begin{document}

\title{Stability of rotation axis of galactic disks in HESTIA suite}

\author{Naira R. Arakelyan}
\affiliation{Astro Space Center of P.N. Lebedev Physical Institute, Moscow, Russia}

\author{Sergey V. Pilipenko}
\email{spilipenko@asc.rssi.ru}
\affiliation{Astro Space Center of P.N. Lebedev Physical Institute, Moscow, Russia}

\author{Stefan Gottl\"{o}ber}
\affiliation{Leibniz-Institut f\"{u}r Astrophysik Potsdam (AIP), An der Sternwarte 16, D-14482 Potsdam, Germany}

\author{Noam I. Libeskind}
\affiliation{Leibniz-Institut f\"{u}r Astrophysik Potsdam (AIP), An der Sternwarte 16, D-14482 Potsdam, Germany}

%\author{Alexander Knebe}
%\affiliation{Departamento de F\'{i}sica Te\'{o}rica, M\'{o}dulo 15, Facultad de Ciencias, Universidad Aut\'{o}noma de Madrid, 28049 Madrid, Spain}
%\affiliation{Centro de Investigaci\'{o}n Avanzada en F\'{i}sica Fundamental (CIAFF), Facultad de Ciencias, Universidad Aut\'{o}noma de Madrid, 28049 Madrid, Spain}
%\affiliation{International Centre for Radio Astronomy Research, University of Western Australia, 35 Stirling Highway, Crawley, Western Australia 6009, Australia}

\author{Gustavo Yepes}
\affiliation{Departamento de F\'{i}sica Te\'{o}rica, M\'{o}dulo 15, Facultad de Ciencias, Universidad Aut\'{o}noma de Madrid, 28049 Madrid, Spain}
\affiliation{Centro de Investigaci\'{o}n Avanzada en F\'{i}sica Fundamental (CIAFF), Facultad de Ciencias, Universidad Aut\'{o}noma de Madrid, 28049 Madrid, Spain}

\author{Yehuda Hoffman}
\affiliation{Racah Institute of Physics, Hebrew University, Jerusalem 91904, Israel}

\date{\today}% It is always \today, today,

\begin{abstract}
We have found that in approximately 1/4 of Milky Way or M31 analogues in the HESTIA suite of simulations there is a significant, $>45^\circ$, change in the axis of rotation orientation of the gaseous disk during the last 6 Gyr. In 6 out of 7 such cases the change is likely caused by the accretion of individual satellites, while in one case such a connection is not obvious. All galaxies that underwent a significant change in rotation axis direction also had angular momentum amplitudes below the sample median 6 Gyr ago. {Our findings support the idea that last 6~Gyr was a quiet phase of evolution in the LG. On the other hand, the mechanism we detect, that satellites can flip galaxy orientation, can explain the difficulties in identifying correlations between the rotation axis and the orientation of large-scale structure elements.}
\end{abstract}

\maketitle

\section{Introduction}
\label{sec:introduction}
During our previous work on the stability of the gravitational potential of MW and M31 analogues from the HESTIA simulation suite \cite{Arakelyan25} we have noticed that the orientation of some of the galaxies is not constant with time. In several realizations the rotation axis moved by more than 90 degrees in the past 6 Gyr (at $z<0.6$). In this work we quantify this change and try to figure out what causes it.

It is widely believed that galaxy rotation originates due to the tidal interactions of proto-galaxies (Tidal Torque Theory, \cite{1969ApJ...155..393P,TTT}). This theory predicts a correlation between the spin of galaxies and the orientation of the elements of the large-scale structure (LSS) \cite{1992ApJ...401..441D,2000ApJ...532L...5L}, and considerable effort has been devoted to studying this effect in observations and simulations \cite{2002MNRAS.332..325P,2002MNRAS.332..339P,2004ApJ...613L..41N,2004ApJ...614L...1L,2006ApJ...640L.111T,2010MNRAS.405..274H,2009IJMPD..18..173S,Libeskind12,Antipova25}. The results are quite complex and somewhat contradictory: the strength of the connection and the direction of the alignment differs for various samples of low-mass and massive galaxies, for DM only and hydrodynamical simulations, for inner or outer halo when only DM is considered. The drift of galaxy orientation with time may be one of the reasons that complicates the analysis of the correlation between the rotation axis and the LSS.

Already in the early simulations of angular momentum growth \cite{1984ApJ...286...38W} it has been observed that individual halos show variations of angular momentum at late times. The authors of \cite{2002ApJ...581..799V} propose that the angular momentum of dark matter halos arises from the cumulative, random acquisition of orbital angular momentum during satellite accretion, rather than solely from early tidal torques. Using N-body simulations and an extended Press-Schechter formalism, the authors model halo spin evolution as a random walk process.  The model reproduces simulation results showing that spin parameters fluctuate significantly over time, increasing sharply during major mergers and declining during gradual accretion. 

Luminous matter concentrate towards halo centers, so it should react differently in comparison with halos as a whole during accretion events. Indeed, \cite{2004ApJ...614L...1L} and \cite{2006ApJ...640L.111T} found that luminous galaxies better retain the memory of initial shear-spin correlations than dark matter halos.

Thus, according to the previous investigations a hypothesis arises that the axis drift we observe in some of the HESTIA simulations can be caused by the accretion of satellites. The accretion of satellites itself has been studied in \cite{2022MNRAS.516.4576D} for three highest resolution realizations of HESTIA. In \cite{2022MNRAS.516.4576D} it was found that the accretion direction is mainly aligned with the axis of slowest collapse of the surrounding material for the satellites that accrete at $z>0.7$. At later times and smaller redshifts this alignment is weaker. Thus, while early accretion may tend to maintain the orientation of the angular momentum with respect to the LSS, later accretion events could randomize the rotation axis.

\section{Large orientation change in some MW and M31 analogues}
\subsection{LG analogues sample}
Simulations from the HESTIA project are made  in a box with $L_{box}=100$ Mpc/h side based on the reconstruction of initial conditions from the measured distribution of velocities of the galaxies around us. They successfully reproduce the large-scale environment of the Local Group (Virgo cluster, local filament, etc.), and the main parameters of the Local Group (LG) itself: the masses of the two main galaxies, their relative distance and velocity. The set of models contains 14 realizations in which other, unmeasured parameters are varied. 13 of these realizations are described in \cite{Libeskind20}, they belong to three groups with identical initial random seeds for the long wave part of the initial conditions, corresponding to $256^3$ mesh in the full box. The group seed is encoded in the first two digits of a simulation code. Within each group simulations have different randomly added small scale structure which is encoded by the second pair of digits.

We use the intermediate resolution runs that have a spatial and mass resolution corresponding to $4096^3$ particles in the box. This resolution is achieved in a 6 Mpc (4 Mpc/h) radius spherical blob around the center of LG, which is free of low resolution particles. These intermediate resolution simulations are available for all 14 realizations. Three from those 14 realizations are also simulated with high resolution corresponding to $8192^3$ particles what allows us to test the effect of resolution. We consider only the last 38 snapshots from the simulations which corresponds to the last six Gyr (redshifts $0<z<0.625$).

All the distances in this Paper are given in physical (not comoving) units. HESTIA assumed the Planck cosmological model with $H_0 = 100h = 67.7$~km/s/Mpc, $\Omega_m=0.31$, $\Omega_\Lambda=0.69$. {Halos and subhalos were identified using the publicly available \texttt{AHF}\footnote{ \texttt{AHF} is publicly available from \url{http://popia.ft.uam.es/AHF/}} halo finder \cite{Knollmann09}.}
Like in \cite{Arakelyan25}, we assign two main roles (MW/M31) in the LG according to relative halo positions in supergalactic coordinates at $z=0$ and not by their masses.

\begin{figure*}
    \centering
    \includegraphics[width=0.99\linewidth]{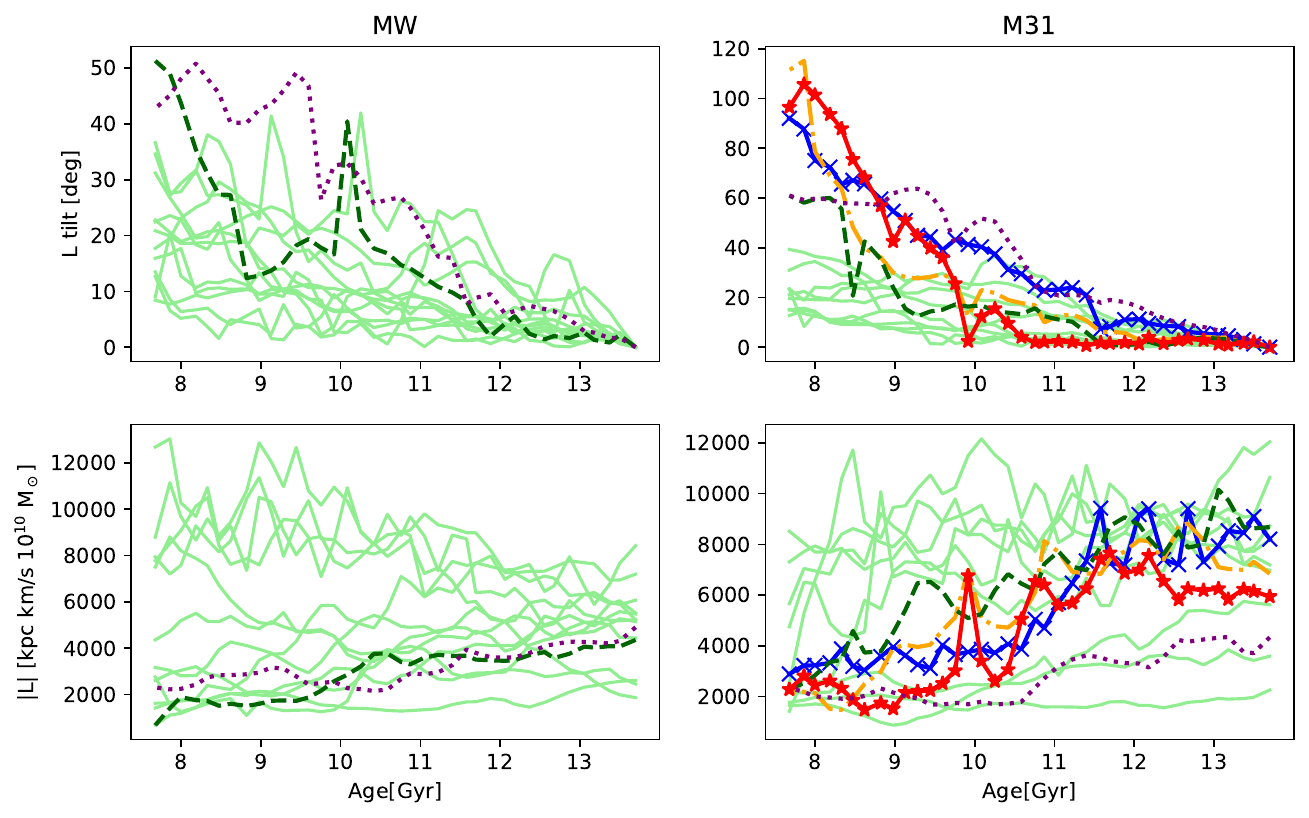}
    \caption{Evolution of the rotation axis of gas within central 20~kpc in MW (left panels) and M31 (right panels) analogues. Top panels: angle between $\vec{L}$ at given time and at $z=0$. Bottom panels: evolution of the amplitude of angular momentum. Realizations in which the angle drifts by more than 45 degrees: 09\_10 (blue solid line with crosses), 09\_16 (orange dot-dashed line), 09\_17 (green dashed line), 09\_18 (red solid line with stars), 37\_16 (purple dotted line). Other realizations are shown with thin light-green lines.}
    \label{fig:L_tilt}
\end{figure*}

\subsection{Axis of rotation}
{HESTIA simulations provide data on particle positions and velocities for three particle types: DM, stars and gas. To define the axis of rotation one must choose particle type and select a region to measure angular momentum. For the former, the choice of gas particles is motivated by the physics of disk galaxies: the gaseous disk is maintained in equilibrium by its rotation. For the comparison, a stellar system may retain its equilibrium due to the angular momenta of individual stars, which may differ significantly from the net angular momentum. It is also known that there is a misalignment of the star and gas axis of rotation in observed and simulated galaxies, see, e.g. \cite{2021ApJS..254...27K,2022MNRAS.514.2031C}, so the choice could presumably affect the rotation axis evolution we want to track.}

{We measure the angular momentum vector $\vec{L}$ of all gas particles (actually, pseudo gas particles defined by the AHF halo finder based on the gas distribution in the AREPO Voronoi cells) found in a sphere around the galaxy center given by AHF (we do not select only gravitationally bound particles). The sphere radius is chosen to be 20~kpc; we varied it from 5 to 40~kpc and verified that our results do not change significantly. 20~kpc is a reasonable size estimate of our Galaxy's disk \cite{2008A&A...487..951K,2010ApJ...718..683C}: the HI disk is much larger in extent  but it is warped and lopsided at large distances. Further we also use 5--40~kpc measurements to determine the causes of rotation axis drift.}

{As an additional check we also measure the orientation of the smallest eigenvector of the gyration tensor $\vec{G}$ of gas particles within the same 20~kpc, defined as
\begin{equation}
    G_{ij} = \frac{1}{N} \sum_{k=1}^{N} x_{i}^k x_{j}^k
\end{equation}
where $N$ is the number of objects, $x_{i}^k$ -- the distance of $k$-th object to the Galactic center along $i$-th coordinate axis. We have found that the smallest eigenvector of the gyration tensor is aligned with the gas $\vec{L}$ within 20$^\circ$ over the last 6 Gyr in all realizations, which shows that the selected gas distribution is really disk-shaped.}

The orientation change of $\vec{L}$ is measured as the angular distance $\theta(t)$ between the angular momentum vector at $z=0$ ($t=t_0$) and at any other given time. We take the vector direction into account, so this angle measures from 0 to 180 degrees:
\begin{equation}
    \theta(t) = arccos((\vec{\ell}(t) , \vec{\ell}(t_0)),\; \vec{\ell}\equiv\vec{L}/|\vec{L}|.
\end{equation}

The evolution of this angle, which we call the ``L tilt'', over the last 6 Gyr is shown in Fig. 1 for all 14 realizations and for both main LG members. We mark all galaxy realizations in which $\vec{L}$ differs by more than 45$^\circ$ from $z=0$ at any epoch within 6 Gyr.

{We chose the critical level of the tilt as 45 degrees from the following considerations. Consider the rotation axis $\vec{L}$ is aligned with some vector $\vec{a}_1$ from an orthogonal basis $\vec{a}_i$, $i=1,2,3$ (e.g. LSS shear tensor eigenvectors), so the scalar product is $(\vec{L},\vec{a}_1)=1$ while $(\vec{L},\vec{a}_i)=0$ for $i=2,3$. If the rotation axis is changed by more than 45 degrees, one can find that the new $\vec{L}'$ is more aligned with some other axis: $(\vec{L}',\vec{a}')>(\vec{L}',\vec{a}_1)$ where $\vec{a}' \perp \vec{a}_1$. In other words, deviations of more than 45 degrees will destroy a previously established matching with LSS elements directions (assuming the LSS itself changes much slower).}

{From Fig. 1 one can see that in the following realizations the tilt larger than 45 degrees is observed on the 6 Gyr interval in M31: 09\_10, 09\_16, 09\_17, 09\_18, 37\_16 and in 09\_17 and 37\_16 it is seen in the MW analogues.}

{After achieving $\vec{L}$ evolution for gas we revisit once again the choice of gas vs. stars. The misalignment between $\vec{L}$ for gas and stars is found to reach up to 50$^\circ$ over 6 Gyr interval, but in all the realizations it drops to $<10^\circ$ at $z=0$. More important for our analysis, changing selection from gas to stars does not change the list of marked galaxies that tilted by more than 45$^\circ$.}

{We also compared the evolution of the total angular momentum inside each galaxy's virial radius for all three components (DM, gas and stars). First, all three virial angular momenta show average misalignment with our $\vec{L}_\mathrm{gas}(20\;\mathrm{kpc})$ of 14--20 degrees over the last 6 Gyr, with maximal misalignment of 80$^\circ$. Second, if we use each of the virial angular momenta to identify large tilt, $>45^\circ$, 9 galaxies show this for $\vec{L}_\mathrm{DM}(R_{vir})$, 9 galaxies for $\vec{L}_\mathrm{gas}(R_{vir})$ and 11 for $\vec{L}_\mathrm{star}(R_{vir})$. Only 3 of them: M31 analogues in 09\_10, 09\_17 and 37\_16 are present in all these large tilt samples, as well as in our main one, based on $\vec{L}_\mathrm{gas}(20\;\mathrm{kpc})$.}

\begin{figure*}
    \centering
    \includegraphics[width=0.99\linewidth]{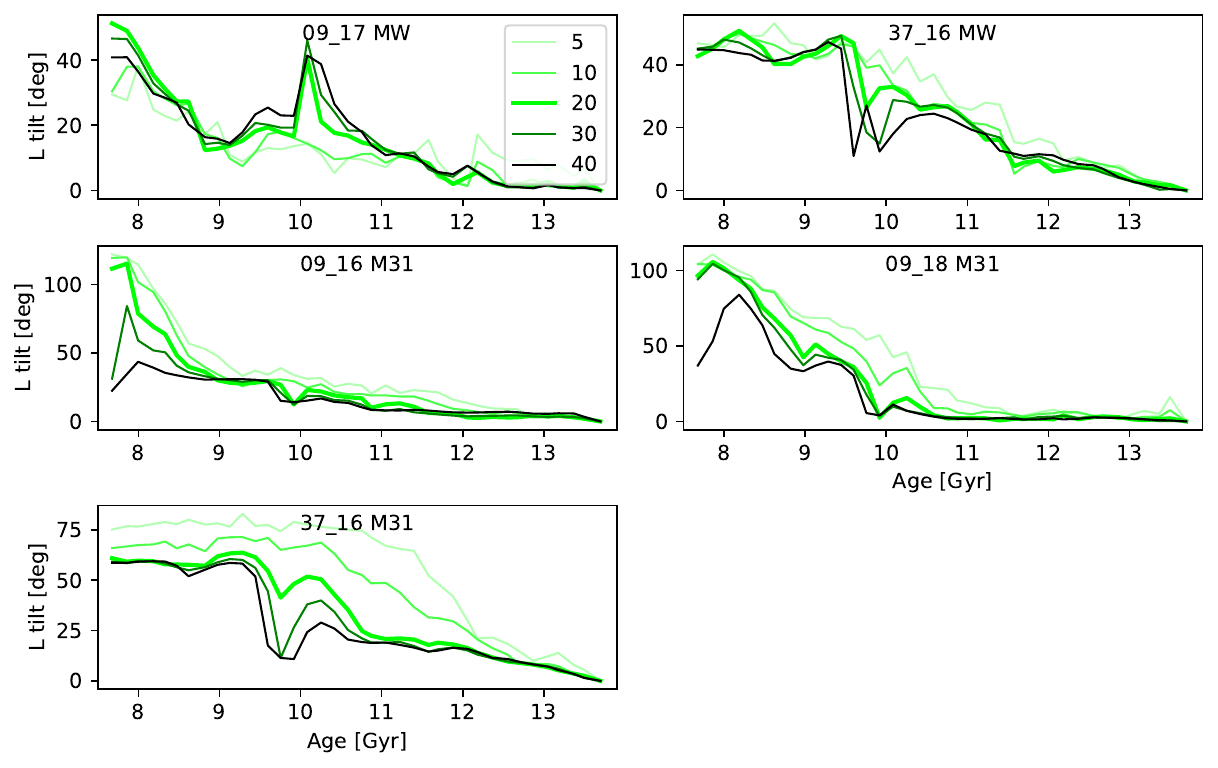}
    \caption{{Evolution of the angular momentum orientation of MW analogues in realizations 09\_17 and 37\_16 and of M31 analogues in realizations 09\_16, 09\_18 and 37\_16. Lines correspond to measurements within spheres of radius r = 5, 10, 20, 30, 40~kpc.}}
    \label{fig:rad}
\end{figure*}

\begin{figure*}
    \centering
    \includegraphics[width=0.99\linewidth]{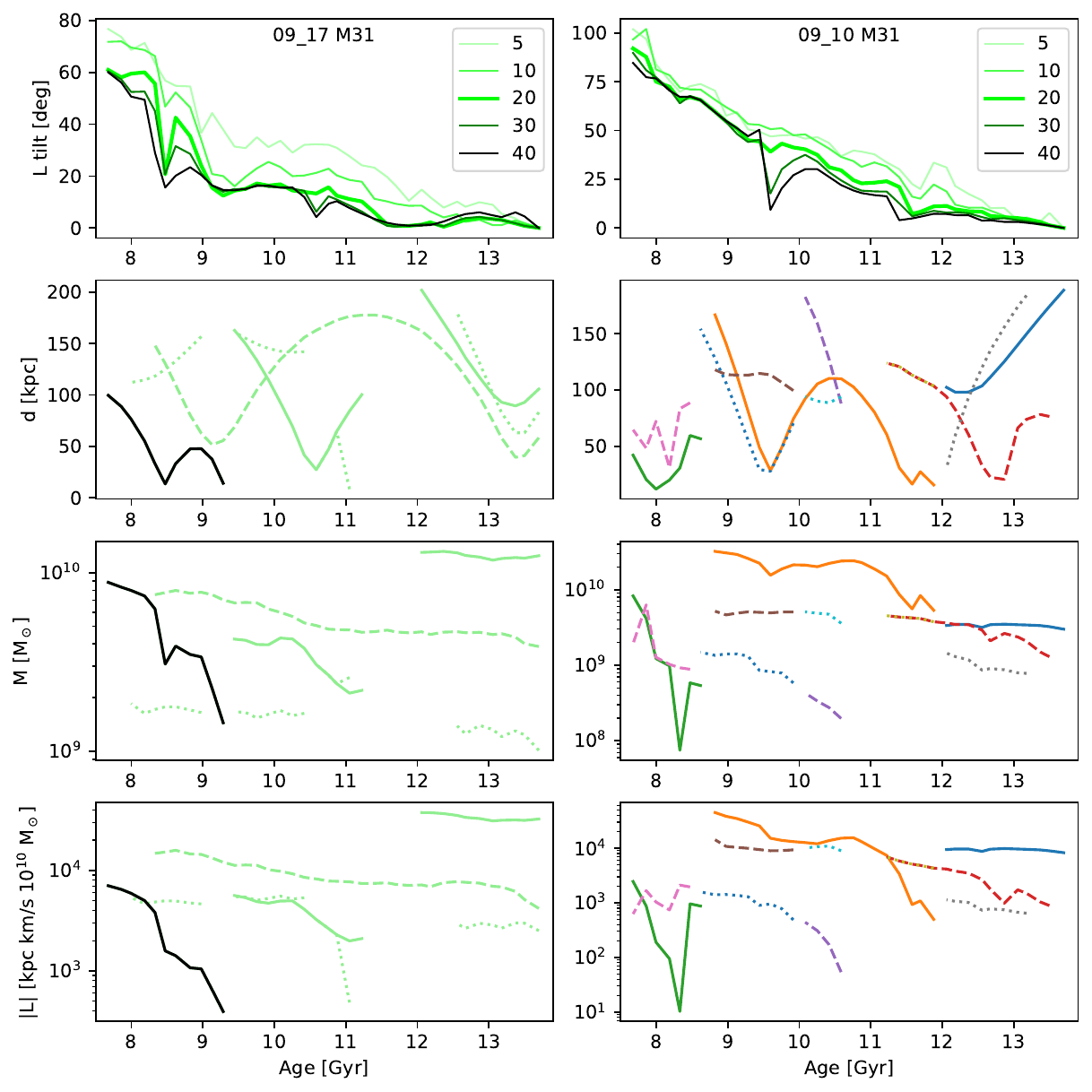}
    \caption{Evolution of the angular momentum orientation of M31 analogues in realizations 09\_17 (left column) and 09\_10 (right column) together with distances, masses of gas and amplitudes of angular momenta of three most massive satellites on each snapshot. Each continuous segment represents one satellite. In the left column, the satellite highlighted in black crosses 40 kpc at ~8.2 Gyr and is directly correlated with the L tilt drop. In the right column there is no single satellite responsible for the change of angular momentum.}
    \label{fig:sat}
\end{figure*}

\begin{figure*}
    \centering
    \includegraphics[width=0.99\linewidth]{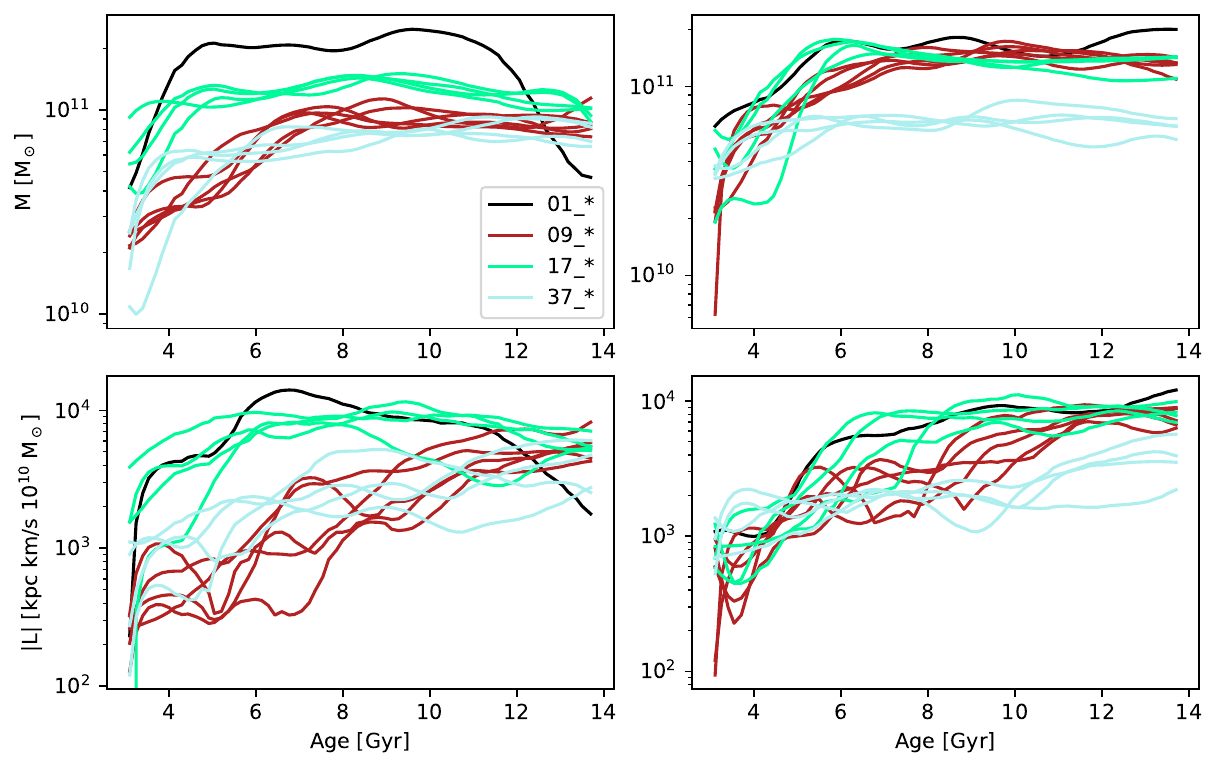}
    \caption{{Long-term evolution of the gas mass (top panels) and angular momentum of gas within 20~kpc from the centers (bottom panels) of two main galaxies in four HESTIA groups of LG analogues. Each group of simulations is shown with its own color. Left panels: MW analogues, right panels: M31 analogues.}}
    \label{fig:LGhist}
\end{figure*}

\section{Causes of orientation change}
While previous work \cite{2002ApJ...581..799V} has shown that satellite accretion changes the angular momentum orientation with time, this may be not the only cause of this change. Accretion of diffuse matter and tidal interaction from the satellites and the surrounding matter (including the second LG member) could also induce changes in the angular momentum. In order to study the origin of the  rotation axis drift, we track gas mass accretion histories (MAHs) of the two main LG members, absolute value of the angular momentum, distance and angular position of the second LG member, distances, masses and angular momenta of the three most massive satellites. We also did the visual inspection of galaxies shape and orientation in 3D using \texttt{gadgetviewer} application.

We have found that all the galaxies that have significantly changed their orientation have one common property: they all had low angular momentum, at least 2 times below the median value, at the time we start our analysis (6 Gyr ago). They all have increased the absolute value of the angular momentum by 1.9--6.4 times from $z=0.625$ to $z=0$. This is seen on the bottom panel of Fig.~\ref{fig:L_tilt} where the angular momentum accretion history (LAH) is shown.

The MAHs of galaxies, measured at several radii, do not show significant changes except the rapid growth when a satellite crosses the corresponding shell. Thus there is no significant accretion of the diffuse material. {The distance between the two main galaxies also changes steadily by no more than 15\% during this interval, which cannot explain rapid and significant $\vec{L}$ changes by tidal effects.}

In 5 out of 7 cases when the axis changes by more than 45 degrees, the analysis of $\vec{L}$ computed in spheres with radius $r=5,10,20,30,40$~kpc shows a gradual change first starting at the largest radius and then propagating to smaller ones. Those 5 cases include realization 37\_16 of MW, and realizations 09\_16, 09\_17, 09\_18, 37\_16 for M31 analogues. The $\vec{L}$ tilt for different radii are shown in Figs.~\ref{fig:rad}, \ref{fig:sat}.  We interpret the propagation from large to small radii as an indication of a satellite with large angular momentum crossing shells while approaching to the center. 

Analysis of distances to the satellites confirms this assumption. An example of realization, in which the angular momentum change is obviously caused by an accreted satellite is shown in the left column of Fig.~\ref{fig:sat}: {the satellite with gas mass $8-9\times10^{9}$~M$\odot$ crosses 40~kpc shell at age of 8.2~Gyr, and gets merged with remnant mass falling to $10^9$~M$\odot$ by 9.2~Gyr. The L tilt at 40~kpc falls from 50 to 20 degrees at 8.2-8.4 Gyr, tilt at 20~kpc falls from 60 to 20 degrees from 8.3 to 9.2 Gyr, while the tilt at 5 and 10~kpc falls more slowly. }

All of the 5 mentioned above cases show that the L tilt change is directly related to the first or second massive satellite approaching to its orbit pericenter. In all these cases the satellite itself carries large orbital angular momentum (larger than the momentum of the galaxy in the beginning of our time interval). The satellites lose their orbital angular momentum and mass during this approach. Their gaseous masses at the beginning of the analysis vary between $4\times10^9$~M$_\odot$ and $10^{10}$~M$_\odot$.  

{From the two remaining cases, in 09\_17 MW analogue there are two satellites infalling to the main galaxy almost simultaneously, one with $M_g\approx2\times10^9$~M$_\odot$ reaching 20~kpc at 8.1 Gyr, and second with $M_g\approx2\times10^9$~M$_\odot$ reaching 20~kpc at 7.8~Gyr. They explain the drop of L tilt from 45 to 20 degrees around age of 8 Gyr. }

The last case, M31 analogue in 09\_10 realization, is the most problematic: it has a very gradual change in the axis direction, almost linearly changing with time, simultaneously on all radii, see right panel of Fig.~\ref{fig:sat}. There are several satellite interaction events, the most prominent one is a $2\times10^{10}$~M$_\odot$ (gas mass) satellite reaching 30 kpc at an age of 9.5 Gyr. But this event causes only a spike in $\vec{L}$ orientation at 30 and 40~kpc, which then returns to the initial trend. {That satellite does not lose its mass and angular momentum during this event, but it does it later, at 11 Gyr during its second pericenter passage. However, the L tilt of the host galaxy is already 25 degrees at that time, so this satellite accretion cannot explain the earlier change from 90 to 25 degrees.}

{It is also important to note that all galaxies with large tilt come from two groups of realizations: 09\_* and 37\_*. The groups differ by the random seed used to fill the data not constrained from the CosmicFlows-2 catalog \cite{2013AJ....146...86T,Libeskind20}. The gas MAHs and LAHs of those groups are shown in Fig.~\ref{fig:LGhist} over 10 Gyr period to capture events before our analyzed last 6 Gyr. 5 out of 7 axis flipping galaxies belong to group 09\_*. All galaxies from this group demonstrate low angular momentum of galactic disks 6 Gyr ago, followed by growth to $z=0$. So in this group the final build-up of angular momentum happens exactly at the period we analyze. In groups 01\_* and 17\_* the main growth of angular momentum happened earlier, 10-6 Gyr ago. Group 37\_*, in which we find also two galaxies with axis flip, shows lower mass and angular momentum of the main galaxies at low redshifts than the other two groups, with less pronounced growth during the last 6 Gyr than 09\_*.}

{It is known that angular momenta of halos and galaxies are more unstable than masses and positions under resolution change in codes like MUSIC and Ginnungagap \cite{Onorbe14,Pilipenko26}, so we also check the effects of resolution on our analysis. HESTIA suite contains 3 realizations computed with the effective resolution 8129$^3$ inside the zoom region, (i.e. all particle masses 8 times smaller than in the 14 realizations we used above). From those 3 realizations, in 5 out of 6 main galaxies their gaseous disk axis of rotation differs from the lower resolution version by no more than 20$^\circ$ during last 6 Gyr. This includes M31 analogue in 09\_18 realization, for which we find large $\vec{L}$ tilt in both 4096$^3$ and 8192$^3$ runs. But in M31 analogue in 37\_11 realization we also see the large tilt ($>60^\circ$) in the high resolution simulation, while in the 4096$^3$ version it is below 40$^\circ$. The difference arises due to a satellite with gas mass $10^{10}$~M$_\odot$ which passes at 100~kpc from galactic center in the lower resolution run, but reaches 20~kpc in the higher resolution run. }

\section{Conclusions}
{Our investigations of the analogues of MW and M31 in the HESTIA suite of simulations have shown that 21 out of the 28 HESTIA  MW and M31 analogues retain the direction of their (gaseous component) spins during the last 6 Gyr. This implies that past 6 Gyr of LG evolution is a quiet phase.}

On the other hand, in 7 out of 28 instances of two main galaxies there was a significant change of the axis of rotation orientation by more than 45 degrees during the last 6 Gyr. We have observed that accretion of satellites with gas mass $>4\times10^{9}$ can significantly change the axis of rotation. This finding may be important to understand the difficulties in identifying the correlation between the axis of rotation and LSS elements orientation. Also it may be important to the analysis of the stability of plain of satellites, like performed in \cite{2025ARep...69.1416P} -- disk axis drift will cause the change of gravitational potential which can distort the satellite orbits.

The galaxies which experienced such a change in their angular momentum direction all had low angular momenta 6 Gyr ago. The change of orientation also was accompanied by the growth of the amplitude of the angular momentum.

While our findings support the idea that satellite accretion is the main cause of the late changes in the angular momentum, in one of the realizations we could not confirm the direct role of satellites: we observe a very gradual, quasi-linear change in the angular momentum orientation, occurring simultaneously at several radii. Analysis of a larger sample of galaxies could shed light on other possible mechanisms  of angular momentum evolution.

\begin{acknowledgments}
The authors thank Alexander Knebe for useful comments.

The authors gratefully acknowledge the Gauss
Centre for Supercomputing e.V. (\url{www.gauss-centre.eu}) for funding
HESTIA project by providing computing time on the GCS Supercomputer
SuperMUC at Leibniz Supercomputing Centre (\url{www.lrz.de})

GY would like to thank the Spanish Agencia Estatal de Investigación of the  Ministerio de Ciencia e Innovaci\'on (AEI/MICINN) for financial support under the project grant PID2024-156100NB-C21.
YH acknowledges the support of the Israeli Science Foundation (grant 1450/24).
\end{acknowledgments}



 \bibliographystyle{elsarticle-num} 
 \bibliography{Grav_pot}

@ARTICLE{Onorbe14,
       author = {{O{\~n}orbe}, Jose and {Garrison-Kimmel}, Shea and {Maller}, Ariyeh H. and {Bullock}, James S. and {Rocha}, Miguel and {Hahn}, Oliver},
        title = "{How to zoom: bias, contamination and Lagrange volumes in multimass cosmological simulations}",
      journal = {\mnras},
         year = 2014,
        month = jan,
       volume = {437},
       number = {2},
        pages = {1894-1908},
          doi = {10.1093/mnras/stt2020},
archivePrefix = {arXiv},
       eprint = {1305.6923},
 primaryClass = {astro-ph.CO},
       adsurl = {https://ui.adsabs.harvard.edu/abs/2014MNRAS.437.1894O}
}

@ARTICLE{2013AJ....146...86T,
       author = {{Tully}, R. Brent and {Courtois}, H{\'e}l{\`e}ne M. and {Dolphin}, Andrew E. and {Fisher}, J. Richard and {H{\'e}raudeau}, Philippe and {Jacobs}, Bradley A. and {Karachentsev}, Igor D. and {Makarov}, Dmitry and {Makarova}, Lidia and {Mitronova}, Sofia and {Rizzi}, Luca and {Shaya}, Edward J. and {Sorce}, Jenny G. and {Wu}, Po-Feng},
        title = "{Cosmicflows-2: The Data}",
      journal = {\aj},
         year = 2013,
        month = oct,
       volume = {146},
       number = {4},
          eid = {86},
        pages = {86},
          doi = {10.1088/0004-6256/146/4/86},
archivePrefix = {arXiv},
       eprint = {1307.7213},
 primaryClass = {astro-ph.CO},
       adsurl = {https://ui.adsabs.harvard.edu/abs/2013AJ....146...86T}
}

@ARTICLE{Pilipenko26,
       author = {{Pilipenko}, Sergey and {Yepes}, Gustavo and {Gottl{\"o}ber}, Stefan and {Knollmann}, Steffen},
        title = "{Ginnungagap {\textemdash} A massively parallel cosmological initial conditions generator}",
      journal = {Astronomy and Computing},
         year = 2026,
        month = apr,
       volume = {55},
          eid = {101082},
        pages = {101082},
          doi = {10.1016/j.ascom.2026.101082},
archivePrefix = {arXiv},
       eprint = {2511.10353},
 primaryClass = {astro-ph.IM},
       adsurl = {https://ui.adsabs.harvard.edu/abs/2026A&C....5501082P}
}

@ARTICLE{2010ApJ...718..683C,
       author = {{Carraro}, Giovanni and {V{\'a}zquez}, Ruben A. and {Costa}, Edgardo and {Perren}, Gabriel and {Moitinho}, Andr{\'e}},
        title = "{The Edge of the Young Galactic Disk}",
      journal = {\apj},
         year = 2010,
        month = aug,
       volume = {718},
       number = {2},
        pages = {683-694},
          doi = {10.1088/0004-637X/718/2/683},
archivePrefix = {arXiv},
       eprint = {1006.1277},
 primaryClass = {astro-ph.GA},
       adsurl = {https://ui.adsabs.harvard.edu/abs/2010ApJ...718..683C}
}

@ARTICLE{2008A&A...487..951K,
       author = {{Kalberla}, P.~M.~W. and {Dedes}, L.},
        title = "{Global properties of the H I distribution in the outer Milky Way. Planar and extra-planar gas}",
      journal = {\aap},
         year = 2008,
        month = sep,
       volume = {487},
       number = {3},
        pages = {951-963},
          doi = {10.1051/0004-6361:20079240},
archivePrefix = {arXiv},
       eprint = {0804.4831},
 primaryClass = {astro-ph},
       adsurl = {https://ui.adsabs.harvard.edu/abs/2008A&A...487..951K}
}

@ARTICLE{2021ApJS..254...27K,
       author = {{Khim}, Donghyeon J. and {Yi}, Sukyoung K. and {Pichon}, Christophe and {Dubois}, Yohan and {Devriendt}, Julien and {Choi}, Hoseung and {Bryant}, Julia J. and {Croom}, Scott M.},
        title = "{Star-Gas Misalignment in Galaxies. II. Origins Found from the Horizon-AGN Simulation}",
      journal = {\apjs},
         year = 2021,
        month = jun,
       volume = {254},
       number = {2},
          eid = {27},
        pages = {27},
          doi = {10.3847/1538-4365/abf043},
archivePrefix = {arXiv},
       eprint = {2012.04659},
 primaryClass = {astro-ph.GA},
       adsurl = {https://ui.adsabs.harvard.edu/abs/2021ApJS..254...27K}
}

@ARTICLE{2022MNRAS.514.2031C,
       author = {{Casanueva}, Catalina I. and {Lagos}, Claudia del P. and {Padilla}, Nelson D. and {Davison}, Thomas A.},
        title = "{The origin of star-gas misalignments in simulated galaxies}",
      journal = {\mnras},
         year = 2022,
        month = aug,
       volume = {514},
       number = {2},
        pages = {2031-2048},
          doi = {10.1093/mnras/stac523},
archivePrefix = {arXiv},
       eprint = {2110.00408},
 primaryClass = {astro-ph.GA},
       adsurl = {https://ui.adsabs.harvard.edu/abs/2022MNRAS.514.2031C}
}

@ARTICLE{2000ApJ...532L...5L,
       author = {{Lee}, Jounghun and {Pen}, Ue-Li},
        title = "{Cosmic Shear from Galaxy Spins}",
      journal = {\apjl},
         year = 2000,
        month = mar,
       volume = {532},
       number = {1},
        pages = {L5-L8},
          doi = {10.1086/312556},
archivePrefix = {arXiv},
       eprint = {astro-ph/9911328},
 primaryClass = {astro-ph},
       adsurl = {https://ui.adsabs.harvard.edu/abs/2000ApJ...532L...5L}
}

@ARTICLE{2025ARep...69.1416P,
       author = {{Pilipenko}, S.~V. and {Arakelyan}, N.~R.},
        title = "{Stability of the Milky Way Satellite Galaxy Plane under the Influence of Neighbors}",
      journal = {Astronomy Reports},
         year = 2025,
        month = dec,
       volume = {69},
       number = {12},
        pages = {1416-1423},
          doi = {10.1134/S1063772926700071},
archivePrefix = {arXiv},
       eprint = {2512.05801},
 primaryClass = {astro-ph.GA},
       adsurl = {https://ui.adsabs.harvard.edu/abs/2025ARep...69.1416P}
}

@ARTICLE{1992ApJ...401..441D,
       author = {{Dubinski}, John},
        title = "{Cosmological Tidal Shear}",
      journal = {\apj},
         year = 1992,
        month = dec,
       volume = {401},
        pages = {441},
          doi = {10.1086/172076},
       adsurl = {https://ui.adsabs.harvard.edu/abs/1992ApJ...401..441D}
}

@ARTICLE{Arakelyan25,
       author = {{Arakelyan}, Naira R. and {Pilipenko}, Sergey V. and {Gottl{\"o}ber}, Stefan and {Libeskind}, Noam I. and {Yepes}, Gustavo and {Hoffman}, Yehuda},
        title = "{Variable gravitational potential of Milky Way analogs in the HESTIA suite}",
      journal = {\prd},
         year = 2025,
        month = aug,
       volume = {112},
       number = {4},
          eid = {043515},
        pages = {043515},
          doi = {10.1103/m2tm-tbm4},
archivePrefix = {arXiv},
       eprint = {2412.18880},
 primaryClass = {astro-ph.GA},
       adsurl = {https://ui.adsabs.harvard.edu/abs/2025PhRvD.112d3515A}
}

@ARTICLE{Antipova25,
       author = {{Antipova}, A.~V. and {Makarov}, D.~I. and {Libeskind}, N.~I. and {Tempel}, E.},
        title = "{Orientation of galaxy spins relative to filaments of the large-scale structure of the Universe}",
      journal = {\pasa},
         year = 2025,
        month = jun,
       volume = {42},
          eid = {e084},
        pages = {e084},
          doi = {10.1017/pasa.2025.10041},
archivePrefix = {arXiv},
       eprint = {2507.07334},
 primaryClass = {astro-ph.GA},
       adsurl = {https://ui.adsabs.harvard.edu/abs/2025PASA...42...84A}
}

@ARTICLE{Libeskind12,
       author = {{Libeskind}, Noam I. and {Hoffman}, Yehuda and {Knebe}, Alexander and {Steinmetz}, Matthias and {Gottl{\"o}ber}, Stefan and {Metuki}, Ofer and {Yepes}, Gustavo},
        title = "{The cosmic web and the orientation of angular momenta}",
      journal = {\mnras},
         year = 2012,
        month = mar,
       volume = {421},
       number = {1},
        pages = {L137-L141},
          doi = {10.1111/j.1745-3933.2012.01222.x},
archivePrefix = {arXiv},
       eprint = {1201.3365},
 primaryClass = {astro-ph.CO},
       adsurl = {https://ui.adsabs.harvard.edu/abs/2012MNRAS.421L.137L}
}

@ARTICLE{2022MNRAS.516.4576D,
       author = {{Dupuy}, Alexandra and {Libeskind}, Noam I. and {Hoffman}, Yehuda and {Courtois}, H{\'e}l{\`e}ne M. and {Gottl{\"o}ber}, Stefan and {Grand}, Robert J.~J. and {Knebe}, Alexander and {Sorce}, Jenny G. and {Tempel}, Elmo and {Tully}, R. Brent and {Vogelsberger}, Mark and {Wang}, Peng},
        title = "{Anisotropic satellite accretion on to the Local Group with HESTIA}",
      journal = {\mnras},
         year = 2022,
        month = nov,
       volume = {516},
       number = {3},
        pages = {4576-4584},
          doi = {10.1093/mnras/stac2486},
archivePrefix = {arXiv},
       eprint = {2208.14648},
 primaryClass = {astro-ph.GA},
       adsurl = {https://ui.adsabs.harvard.edu/abs/2022MNRAS.516.4576D}
}

@ARTICLE{Knollmann09,
       author = {{Knollmann}, Steffen R. and {Knebe}, Alexander},
        title = "{AHF: Amiga's Halo Finder}",
      journal = {\apjs},
         year = 2009,
        month = jun,
       volume = {182},
       number = {2},
        pages = {608-624},
          doi = {10.1088/0067-0049/182/2/608},
archivePrefix = {arXiv},
       eprint = {0904.3662},
 primaryClass = {astro-ph.CO},
       adsurl = {https://ui.adsabs.harvard.edu/abs/2009ApJS..182..608K}
}

@ARTICLE{Libeskind20,
       author = {{Libeskind}, Noam I. and {Carlesi}, Edoardo and {Grand}, Robert J.~J. and {Khalatyan}, Arman and {Knebe}, Alexander and {Pakmor}, Ruediger and {Pilipenko}, Sergey and {Pawlowski}, Marcel S. and {Sparre}, Martin and {Tempel}, Elmo and {Wang}, Peng and {Courtois}, H{\'e}l{\`e}ne M. and {Gottl{\"o}ber}, Stefan and {Hoffman}, Yehuda and {Minchev}, Ivan and {Pfrommer}, Christoph and {Sorce}, Jenny G. and {Springel}, Volker and {Steinmetz}, Matthias and {Tully}, R. Brent and {Vogelsberger}, Mark and {Yepes}, Gustavo},
        title = "{The HESTIA project: simulations of the Local Group}",
      journal = {\mnras},
         year = 2020,
        month = oct,
       volume = {498},
       number = {2},
        pages = {2968-2983},
          doi = {10.1093/mnras/staa2541},
archivePrefix = {arXiv},
       eprint = {2008.04926},
 primaryClass = {astro-ph.GA},
       adsurl = {https://ui.adsabs.harvard.edu/abs/2020MNRAS.498.2968L}
}

@ARTICLE{1969ApJ...155..393P,
       author = {{Peebles}, P.~J.~E.},
        title = "{Origin of the Angular Momentum of Galaxies}",
      journal = {\apj},
         year = 1969,
        month = feb,
       volume = {155},
        pages = {393},
          doi = {10.1086/149876},
       adsurl = {https://ui.adsabs.harvard.edu/abs/1969ApJ...155..393P}
}

@ARTICLE{1984ApJ...286...38W,
       author = {{White}, S.~D.~M.},
        title = "{Angular momentum growth in protogalaxies}",
      journal = {\apj},
         year = 1984,
        month = nov,
       volume = {286},
        pages = {38-41},
          doi = {10.1086/162573},
       adsurl = {https://ui.adsabs.harvard.edu/abs/1984ApJ...286...38W}
}

@ARTICLE{2002MNRAS.332..325P,
       author = {{Porciani}, Cristiano and {Dekel}, Avishai and {Hoffman}, Yehuda},
        title = "{Testing tidal-torque theory - I. Spin amplitude and direction}",
      journal = {\mnras},
         year = 2002,
        month = may,
       volume = {332},
       number = {2},
        pages = {325-338},
          doi = {10.1046/j.1365-8711.2002.05305.x},
archivePrefix = {arXiv},
       eprint = {astro-ph/0105123},
 primaryClass = {astro-ph},
       adsurl = {https://ui.adsabs.harvard.edu/abs/2002MNRAS.332..325P}
}

@ARTICLE{2002MNRAS.332..339P,
       author = {{Porciani}, Cristiano and {Dekel}, Avishai and {Hoffman}, Yehuda},
        title = "{Testing tidal-torque theory - II. Alignment of inertia and shear and the characteristics of protohaloes}",
      journal = {\mnras},
         year = 2002,
        month = may,
       volume = {332},
       number = {2},
        pages = {339-351},
          doi = {10.1046/j.1365-8711.2002.05306.x},
archivePrefix = {arXiv},
       eprint = {astro-ph/0105165},
 primaryClass = {astro-ph},
       adsurl = {https://ui.adsabs.harvard.edu/abs/2002MNRAS.332..339P}
}

@ARTICLE{2002ApJ...581..799V,
       author = {{Vitvitska}, Maya and {Klypin}, Anatoly A. and {Kravtsov}, Andrey V. and {Wechsler}, Risa H. and {Primack}, Joel R. and {Bullock}, James S.},
        title = "{The Origin of Angular Momentum in Dark Matter Halos}",
      journal = {\apj},
         year = 2002,
        month = dec,
       volume = {581},
       number = {2},
        pages = {799-809},
          doi = {10.1086/344361},
archivePrefix = {arXiv},
       eprint = {astro-ph/0105349},
 primaryClass = {astro-ph},
       adsurl = {https://ui.adsabs.harvard.edu/abs/2002ApJ...581..799V}
}

@ARTICLE{2004ApJ...613L..41N,
       author = {{Navarro}, Julio F. and {Abadi}, Mario G. and {Steinmetz}, Matthias},
        title = "{Tidal Torques and the Orientation of Nearby Disk Galaxies}",
      journal = {\apjl},
         year = 2004,
        month = sep,
       volume = {613},
       number = {1},
        pages = {L41-L44},
          doi = {10.1086/424902},
archivePrefix = {arXiv},
       eprint = {astro-ph/0405429},
 primaryClass = {astro-ph},
       adsurl = {https://ui.adsabs.harvard.edu/abs/2004ApJ...613L..41N}
}

@ARTICLE{2004ApJ...614L...1L,
       author = {{Lee}, Jounghun},
        title = "{The Intrinsic Inclination of Galaxies Embedded in Cosmic Sheets and Its Cosmological Implications: An Analytic Calculation}",
      journal = {\apjl},
         year = 2004,
        month = oct,
       volume = {614},
       number = {1},
        pages = {L1-L4},
          doi = {10.1086/425326},
archivePrefix = {arXiv},
       eprint = {astro-ph/0408251},
 primaryClass = {astro-ph},
       adsurl = {https://ui.adsabs.harvard.edu/abs/2004ApJ...614L...1L}
}

@ARTICLE{2006ApJ...640L.111T,
       author = {{Trujillo}, Ignacio and {Carretero}, Conrado and {Patiri}, Santiago G.},
        title = "{Detection of the Effect of Cosmological Large-Scale Structure on the Orientation of Galaxies}",
      journal = {\apjl},
         year = 2006,
        month = apr,
       volume = {640},
       number = {2},
        pages = {L111-L114},
          doi = {10.1086/503548},
archivePrefix = {arXiv},
       eprint = {astro-ph/0511680},
 primaryClass = {astro-ph},
       adsurl = {https://ui.adsabs.harvard.edu/abs/2006ApJ...640L.111T}
}

@ARTICLE{2010MNRAS.405..274H,
       author = {{Hahn}, Oliver and {Teyssier}, Romain and {Carollo}, C. Marcella},
        title = "{The large-scale orientations of disc galaxies}",
      journal = {\mnras},
         year = 2010,
        month = jun,
       volume = {405},
       number = {1},
        pages = {274-290},
          doi = {10.1111/j.1365-2966.2010.16494.x},
archivePrefix = {arXiv},
       eprint = {1002.1964},
 primaryClass = {astro-ph.CO},
       adsurl = {https://ui.adsabs.harvard.edu/abs/2010MNRAS.405..274H}
}

@ARTICLE{2009IJMPD..18..173S,
       author = {{Sch{\"a}fer}, Bj{\"o}rn Malte},
        title = "{Galactic Angular Momenta and Angular Momentum Correlations in the Cosmological Large-Scale Structure}",
      journal = {International Journal of Modern Physics D},
         year = 2009,
        month = jan,
       volume = {18},
       number = {2},
        pages = {173-222},
          doi = {10.1142/S0218271809014388},
archivePrefix = {arXiv},
       eprint = {0808.0203},
 primaryClass = {astro-ph},
       adsurl = {https://ui.adsabs.harvard.edu/abs/2009IJMPD..18..173S}
}

@ARTICLE{TTT,
       author = {{Doroshkevich}, A.~G.},
        title = "{Spatial structure of perturbations and origin of galactic rotation in fluctuation theory}",
      journal = {Astrophysics},
         year = 1970,
        month = oct,
       volume = {6},
       number = {4},
        pages = {320-330},
          doi = {10.1007/BF01001625},
       adsurl = {https://ui.adsabs.harvard.edu/abs/1970Ap......6..320D}
}





% \end{thebibliography}
\end{document}